\documentclass[preprint,aps,superscript,address,showkeys,pre]{revtex4}
\usepackage{amsmath,amssymb}
\usepackage{epsfig}
\usepackage{enumerate}
\usepackage{graphics}
\usepackage{graphicx}
\usepackage{float}
\usepackage{dcolumn}
\begin{document}

\title{Enhanced signal propagation in a network with unidirectional and random couplings}

\author{S.~Rajamani}
\email{rajeebard@gmail.com}
\author{S.~Rajasekar$^{**,}$}
\email{rajasekar@cnld.bdu.ac.in}
\affiliation{School of Physics, Bharathidasan University, Tiruchirapalli 620 024, Tamilnadu, India}

\begin{abstract}

We investigate the effect of unidirectional regular and random couplings of units in a network on stochastic resonance. For simplicity we choose the units as Bellows map with bistability. In a regular network we apply a weak periodic signal and noise to first unit only. Above certain coupling strength undamped and enhanced signal propagation takes place. Resonance occurs at the same value of noise intensity in all units. The response amplitude displays sigmoidal function type variation with unit number. When all the units are driven by the weak periodic force and noise oscillatory variation of response amplitude with unit number occurs. In the network with random coupling all the units are subjected to periodic force and noise. In this case the value of noise intensity at which resonance occurs and the corresponding value of average response amplitude increases nonlinearly with the number of units wired. Numerical simulation with multiple couplings shows that in both type of networks single coupling is sufficient to realize a great improvement in stochastic resonance and signal propagation.

\keywords{network, signal propagation, stochastic resonance, Bellows map. \\ $^{**}$Corresponding author}

\end{abstract}
\maketitle

\newpage

\section{Introduction}
Stochastic resonance (SR) is a noise-induced nonlinear phenomenon in which an appropriate weak signal evokes the best correlation between a weak periodic signal and the output signal of a bistable system. In a recent years, investigation of features of SR in coupled and networks of nonlinear systems received a great deal of interest. This is primarily because of its constructive role in enhancement of weak signal detection and transmission. In coupled arrays it has been shown that the spatiality may broaden the scope of SR {\cite{r1,r2}}. In a network system this noise-induced resonance can be enhanced by a coupling term and this phenomenon is termed as enhanced SR {\cite{r1}}. In Hodgkin-Huxley neuronal networks increase in the network randomness is found to increase the temporal coherence and spatial synchronization {\cite{r3}}. A resonance-like effect depending on the number of systems, whereby an optimal number of systems leading to the maximum overall coherence has been reported {\cite{r4}}. Resonance is found in certain networks where all the systems are driven by noise while only one system is subjected to a weak periodic signal {\cite{r5}}. Doubly SR consisting of a noise-induced phase transition has been observed in a conventional nonlinear lattice of coupled overdamped oscillators {\cite{r6}}. Double resonance peaks were found to occur in the Ising model network driven by an oscillating magnetic field {\cite{r7}}. Improvement of signal-to-noise ratio is possible in an uncoupled parallel array of bistable system subjected to a common noise {\cite{r8}}. SR has been studied in coupled threshold elements {\cite{r9}}, Ising model {\cite{r10}}, Barabasi-Albert network {\cite{r11}}, arrays of comparators {\cite{r12}} and in a B\"ar model {\cite{r13}}.

Investigation of features of different types of couplings and connectivity topology with reference to various nonlinear phenomena in networks is very important because they carry information from one unit to another and act as a source of amplification of information. This would help us to learn and understand the role of connectivity topology in networks. In this connection, in the context of SR, in most of the studies on network systems, noise and a weak periodic signal are applied to all the systems. Moreover, regular and random single and multiple diffusive couplings are considered. It has been shown that a small fraction of long-range coupling is sufficient to have a great improvement in SR and synchronization {\cite{r14}}. Perc {\cite{r15}} reported that enhanced noise-induced resonance can occur only for intermediate coupling in a small-world network. Self-organized phase-shifts are noticed between large degree and small degree nodes when the coupling terms are weighted according to degree of nodes {\cite{r16}}. Perc and his co-workers analyzed SR in a network system consisting of bistable oscillators in which all the oscillators are subjected to noise, each oscillator is connected to $m$ neighborhood oscillators (diffusive coupling) and a weak periodic signal is applied to one of the oscillators only {\cite{r5,r17}}. Effect of connectivity as a function of the location of oscillators and the amplitude of external forcing is also analyzed {\cite{r18}}. Wu et al {\cite{r19}} shown a monotonic increasing of spectral amplification factor with the coupling constant when the coupling is considered as adaptive.

It is of great significance to introduce both weak input signal and weak noise to only one unit of a network and also consider a simple connectivity, for example, unidirectional linear coupling, and investigate the signal propagation through the units. Such a setup has applications in digital sonar arrays, networks of sensory neurons, analog to digital converters where arrays are useful for the transmission of a noisy periodic input signal and the performance is assessed by the signal-to-noise ratio in the frequency domain. For chaos synchronization and communication in coupled lasers often in the transmitter-receiver setup the transmitter laser is unidirectionally coupled to a receiver laser {\cite{r20}}. One-way coupling was shown to improve the performance of flux-gate magnetometers {\cite{r21}} and enhance the high-frequency induced resonance {\cite{r22}}. Very recently, in an electronic two-dimensional array of bistable oscillators with one-way coupling, soliton-like waves are found to propagate in different directions with different speeds {\cite{r23}}. 

Our prime goal in the present work is to investigate the impact of unidirectional regular connection and random connection with linear coupling term on SR in a network with each unit or site represented by a bistable discrete Bellows map. This map possesses the basic requirements to display noise-induced resonance. We have chosen a discrete equation rather than a continuous time evolution equation, as the system of a unit, mainly because the former requires relatively very less computational time and resources. Often discrete maps served as a convenient models for discovering new phenomena and identifying their features. The results of our study, in general, can be realized in networks of other bistable as well as excitable systems with same connectivity topology. 

There are four sections in this paper. In Sec.II we consider a network of Bellows map with the weak input periodic signal and Gaussian white noise applied to first unit only with one-way coupling. The first system is uncoupled. The coupling term is linear. We consider single as well as multiple connections. There are several interesting common and different features associated with the various units. For coupling strength above a critical value undamped signal propagation occurs. In this case all the units exhibit SR and the values of noise intensity $D$ at which the response amplitude $Q$ becomes maximum in various units are the same. However, for a fixed noise intensity $Q$ increases with the unit number $i$ and then it becomes a constant. At resonance periodic switching between two states of the map takes place and the units are unsynchronized. Then we demonstrate the influence of unidirectional multiple couplings, that is, coupling of $i$th unit with $i+1,\,i+2,\,...,\ i+m$ units. We show that one and two couplings ($m=1,2$) are more effective than higher number of couplings. Also we study the case of all the units subjected to weak periodic signal and noise with one-way coupling. For sufficiently large values of coupling strength and for a range of values of noise intensity we observe oscillatory variation of $Q$ with the unit number $i$. This behaviour is pronounced for intermediate values of noise intensity. Section III is devoted to the study of unidirectional but random coupling. We consider the cases of fraction of total units ($25\%,\,50\%,\,75\%$ and $100\%$ of total units) coupled to randomly chosen units. $D_{{\mathrm{max}}}$, the value of $D$ at which average response amplitude $\langle Q \rangle$ of the network becomes maximum and $\langle Q \rangle_{{\mathrm{max}}}$, the maximum value of $\langle Q \rangle $ at resonance, increases with increase in the number of units connected. For a fixed noise intensity $\langle Q \rangle $ exhibits sigmoidal function type variation with the total number of units. Further, $\langle Q \rangle $ increases with increase in multiple number of couplings, however, the increment is small. Finally, we present summary of the results in Sec.IV.

\section{Stochastic resonance in an one-way coupled regular network}
In this section we explore the features of SR in a one-way coupled regular network with $N$ units and $m$ couplings. We consider the cases of periodic signal and noise applied to first unit only and to all the units.
\subsection{Description of the network model}
\begin{figure}[t]
\begin{center}
\epsfig{figure=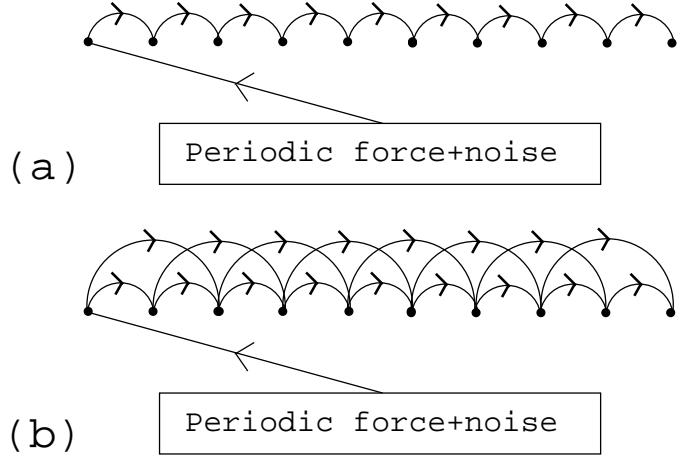, width=0.55\columnwidth}
\end{center}
\caption{Examples of one-way coupled networks with $10$ units. The dynamics of the first unit is independent of the other units and this is alone driven by both weak periodic signal and noise. (a) $i$th unit ($i \ne 1$) is linearly coupled to ($i-1$)th unit. The arrow mark indicates that the output of $i$th unit is fed to ($i+1$)th unit only through the linear coupling term. (b) $i$th unit ($i>2$) is linearly coupled to both ($i-1$)th and ($i-2$)th sites while the second site ($i=2)$ is coupled to the first site only. }
\label{fig1}
\end{figure}

The network consists of $N$ units. The first unit is uncoupled and is alone driven by both weak input periodic signal and noise. The interaction and its range $m$ both are along unidirection. The coupling term is linear and the system representing each unit is the Bellows map {\cite{r24,r25,r26}}. This type of situation is realized in many real-life systems {\cite{r5,r27,r28,r29}}. Figure {\ref{fig1}} depicts the above network topology. 
The dynamics of the network is described by
\begin{subequations}
 \label{eq1}
\begin{eqnarray}
  x_{n+1}^{(1)} 
     & = & \frac{r x_n^{(1)} }{ 1 + \left( x_n^{(1)} \right)^b}
               + f \cos \, \omega n + \sqrt{D} \xi (n), \\
  x_{n+1}^{(i)} 
     & = & \frac{r x_n^{(i)} }{ 1 + \left( x_n^{(i)} \right)^b}
               + \frac{\delta}{M} \sum_{j=1}^m x_n^{(i-j)} ,  \quad 
               i=2,3,...,N, \;\; i-j \ge 1       
\end{eqnarray}
where 
\begin{eqnarray}
  M = \begin{cases}  i-1, & {\mathrm{if}} \; i \le m \\
              m, &  {\mathrm{if}} \; i > m.
       \end{cases}
\end{eqnarray}
\end{subequations}
In Eqs.~(\ref{eq1}) the number of couplings is $m$. For simplicity we choose $N \gg m$. $\xi(n)$ is a Gaussian white noise with the statistical properties $\langle \xi (n) \rangle =0$, $\langle \xi (n) \xi(n') \rangle = \delta(n-n')$, $D$ is noise intensity. We choose the value of $b$ in the Bellows map as $2$. The Bellows map $x_{n+1} =rx_n/(1+x_n^2)$ has only one fixed point $x^* =0$ for $0 < r \le 1$. For $r>1$, it has three fixed points $x^*=0$ (unstable) and $x^*_{\pm} = \pm \sqrt{r-1}$ (stable). The map has bistable state for $r>1$. In order realize SR in the bistable case we drive the map by a weak periodic signal $f \cos \omega n$ and the Gaussian white noise. We fix the values of the parameters in the network as $r=2$, $b=2$, $f=0.3$, $\omega=0.1$, $N=400$ and vary the noise intensity and the coupling strength $\delta$. The values of $f$ is below the subthreshold, that is, in the absence of noise the periodic force alone is unable to induce transition between the two stable fixed points. 

\subsection{Stochastic resonance and signal propagation}

To characterize the noise-induced SR we numerically compute the response amplitude $Q_i$ at the input signal frequency $\omega$. $Q_i$ is given by 
\begin{subequations}
 \label{eq2}
\begin{eqnarray}
   Q_i & = & \frac{\sqrt{Q_{i,C}^2+Q_{i,S}^2}}{f}, \\
   Q_{i,{\mathrm{C}}} & = & \frac{2}{Tt} \sum_{n=1}^{Tt} x_n^{(i)} \cos \,\omega n, \\
   Q_{i,{\mathrm{S}}} & = & \frac{2}{Tt} \sum_{n=1}^{Tt} x_n^{(i)} \sin \,\omega n,
\end{eqnarray}
\end{subequations}
where $t=2 \pi/\omega$ is the period of the input signal $f \cos \omega n$ and $T$ is chosen as $1000$. $Q_i$ is often used as a measure for SR {\cite{r5,r15,r18}}.

Let us point out the occurrence of SR in the first unit of the network the dynamics of which is independent of the other units. As $D$ increases from a small value the response amplitude $Q_1$ increases, reaches a maximum value at $d=D_{{\mathrm{max}}}=0.175$ and decreases with further increase in $D$. For $D \ll D_{{\mathrm{max}}}$ the iterated values rarely switches between the two stable fixed points. At $D=0.175$ the iteration plot, $x_n^{(1)}$ versus $n$, shows almost periodic switching between the regions $x>0$ and $x<0$. Numerically computed mean residence times in the regions $x<0$ and $x>0$ are $\approx T/2$. For $D \gg D_{{\mathrm{max}}}$ erratic switching between these two regions takes place.

\begin{figure}[t]
\begin{center}
\epsfig{figure=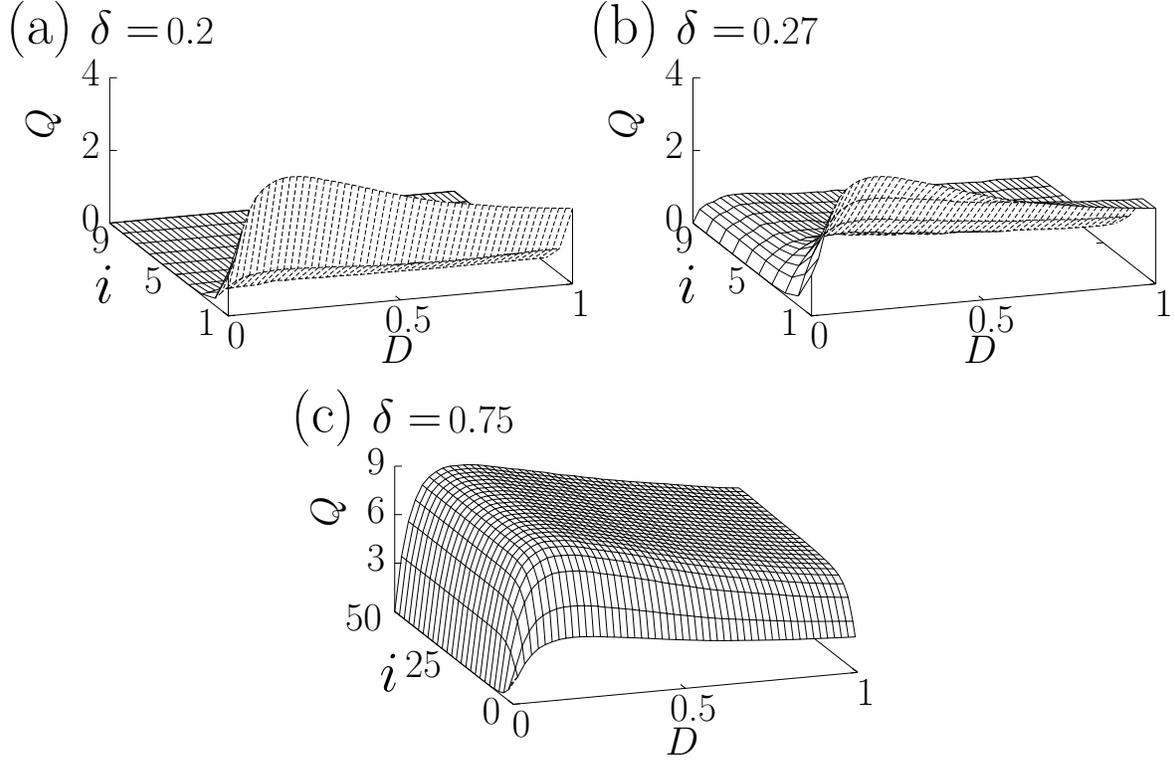, width=0.95\columnwidth}
\end{center}
\caption{Three-dimensional plots of variation of the response amplitude $Q$ as a function of unit number $i$ and noise intensity $D$ for three values of the coupling strength $\delta$ for the network system given by Eqs.~({\ref{eq1}}). The values of the parameters are $r=2$, $b=2$, $f=0.3$, $\omega=0.1$, $N=400$ and $m=1$.}
\label{fig2}
\end{figure}

\begin{figure}[t]
\begin{center}
\epsfig{figure=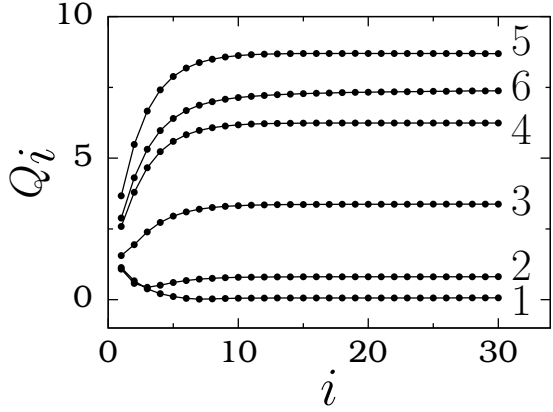, width=0.45\columnwidth}
\end{center}
\caption{Variation of the response amplitude with the unit number $i$ for six fixed values of the noise intensity $D$ with $\delta=0.75$ and $m=1$. The values of $D$ for the curves $1-6$ are $0.002$, $0.005$, $0.02$, $0.05$, $0.175$ and $0.5$ respectively. }
\label{fig3}
\end{figure}

\begin{figure}[t]
\begin{center}
\epsfig{figure=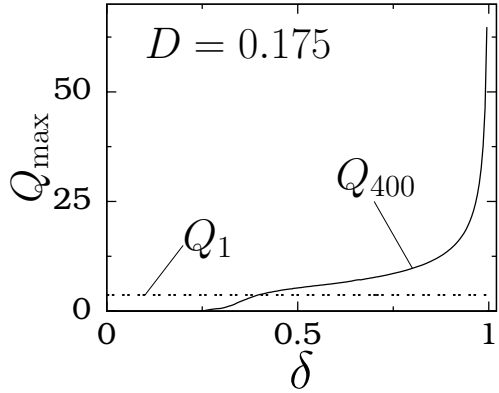, width=0.4\columnwidth}
\end{center}
\caption{$Q_{{\mathrm{max}}}$ of the last unit with the coupling strength $\delta$. The dashed line represents the value of $Q_1$ which is independent of $\delta$.}
\label{fig4}
\end{figure}

We perform extensive numerical simulation for the network system (\ref{eq1}), varying the parameters $\delta$ and $D$. We restrict ourselves to $\delta \in [0,1]$ and $D \in [0,1]$. First we report the resonance dynamics with $m=1$ corresponding to the network topology shown in Fig.~{\ref{fig1}}(a). Figure {\ref{fig2}} presents the effect of coupling strength $\delta $ and $D$ on the response amplitude of the various units. For small values of $\delta$ when $D$ is varied typical SR occurs only in the first few units and in the other units $Q_i=0$ for the entire range of $D$. This is shown in Fig.~{\ref{fig2}}(a). 
In this case for the units far from the first unit $x_n^{(i)}$ becomes a constant after a transient evolution. When $Q_i \ne 0$ the evolution has not settled on to a steady state value. We note that in a random network of Ising model {\cite{r7}} $Q=0$ and $Q\ne 0$ imply paramagnetic and ferromagnetic phases. When $\delta>0.237$ all the units display SR and moreover resonance occurs at the same value of $D$ in all the units. For $0.237 < \delta<0.4$ the response amplitude of the last unit is less than $Q_1$. In Fig.~{\ref{fig2}}(b) for $\delta = 0.27$ $Q_i$ decreases with the unit number $i$. Undamped signal propagation takes place, however, there is no enhancement of the output signal (at the frequency $\omega$) of the last unit. That is, $Q$ of the last unit is $<1$. For $\delta>0.4$ $Q_{{\mathrm{max}}}$ increases and becomes a constant with the unit number $i$. Further, $Q_{{\mathrm{max}}}$ of $i$th unit ($i \ne 1$) is greater than that of the first unit. An example for this is shown in Fig.~{\ref{fig2}}(c) where $\delta=0.75$. For clarity in Fig.~{\ref{fig3}} we plot $Q_i$ versus $i$ for several fixed values of $D$. For each fixed  value of $D$, $Q_i$ displays sigmoidal type variation with $i$. In Fig.~{\ref{fig3}} for $D=0.002$ and $0.005$ we observe $Q_i<Q_1$ while $Q_i>Q_1$ for other values of $D$ .
In the numerical simulation for $\delta>0.4$ we find enhanced undamped signal propagation (that is, $Q_{400}>Q_1$) except for very small values of $D$. Though the weak periodic signal and noise is applied to the first unit alone the linear coupling of it to the second unit is able to stimulate it to exhibit SR and this process propagates to the successive units through simple unidirectional coupling. Figure {\ref{fig4}} displays the variation of $Q_{{\mathrm{max}}}$ of the last unit with the coupling strength $\delta$. It varies rapidly for values of $\delta$ near $1$. For $\delta > 1$ the iterations $x_n^{(1)}$ diverge. 
\begin{figure}[!h]
\begin{center}
\epsfig{figure=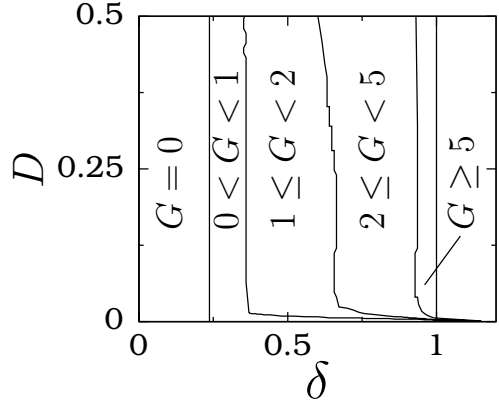, width=0.4\columnwidth}
\end{center}
\caption{Regions of various ranges of gain factor $G$ in the $(\delta,D$) parameter space of the network system given by Eqs.~(\ref{eq1}) with $m=1$. The motion is unbounded for $\delta>1$.}
\label{fig5}
\end{figure}

We define the gain factor $G=Q_N/Q_1$ with $N=400$. We numerically compute it for values of $D$ and $\delta$ in the interval $[0,1]$. Regions in $(\delta,D$) parameter space with $5$ ranges of $G$, namely, $=0$, $<1$, $[1,2]$, $[2,5]$ and $>5$ are depicted in Fig.~{\ref{fig5}}. The coupling strength has a strong influence on $G$. The gain factor $G$ is $0$ for small values of $\delta$ for the entire range of noise intensity considered. On the other hand, it increases with increase in $\delta$ for fixed values of $D$. 

\begin{figure}[t]
\begin{center}
\epsfig{figure=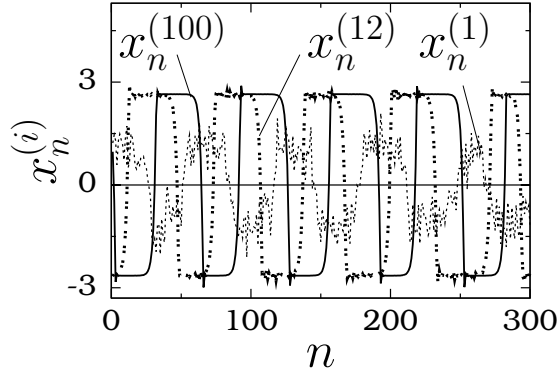, width=0.45\columnwidth}
\end{center}
\caption{$x_n^{(i)}$ versus $n$ for the units $i=1$, $2$ and $100$ of the network (\ref{eq1}) with $m=1$, $\delta=0.75$ and $D=0.175$ at which the response amplitude becomes a maximum. $x_n^{(i)}$ exhibits periodic switching between the regions $x>0$ and $x<0$. }
\label{fig6}
\end{figure}

Figure {\ref{fig6}} presents an interesting result. For $\delta >0.237$ for each fixed value of $D$ the number of oscillations of $x_n^{(i)}$ in the regions $x<0$ ($x>0$) before switching to $x>0$ ($x<0$) decreases with increase in the unit number $i$. This is clearly evident in Fig.~{\ref{fig6}} where $x_n^{(i)}$ versus $n$ is plotted for $i=1$, $12$ and $100$ for $\delta=0.75$ and $D=0.175$. For large $i$ the coupling term weakens the oscillation in the regions $x<0$ and $x>0$ and the output signal appears as a rectangular pulse. That is, the one-way coupling with appropriate strength gives rise undamped propagation of signal in the form of rectangular pulse. The Fourier series of such a signal will contain frequencies $l \omega$ where $l=1,2,...$ with decaying amplitudes. Figure {\ref{fig6}} corresponds to the case at which SR occurs. $x_n^{(i)}$ exhibits almost periodic switching between the two regions $x<0$ and $x>0$. We note that $x_n^{(i)}$'s are not synchronized in the sense that switching of the various units from one region to another region not takes place at almost same value of $n$. We numerically calculate the mean residence time of each unit in the regions $x<0$ and $x>0$. At resonance mean residence times of all the units are found to be $T/2$ where $T=2 \pi / \omega$. 

\begin{figure}[!h]
\begin{center}
\epsfig{figure=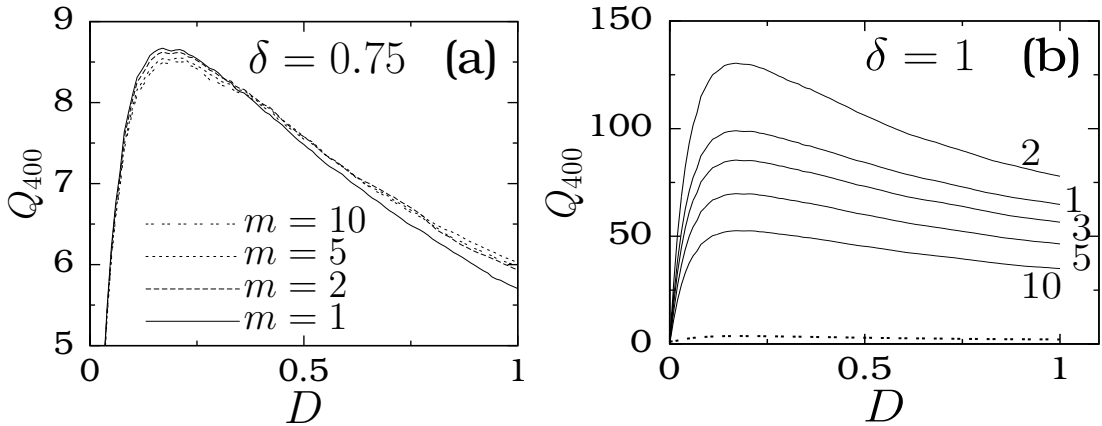, width=0.9\columnwidth}
\end{center}
\caption{The effect of multiple coupling on the response amplitude of the last unit of the network system given by Eqs.~({\ref{eq1}}) for two values of the coupling constant $\delta$. In the subplot (b) the number marked to each curve is the value of $m$ (number of couplings). The dashed curve is the response amplitude $Q_1$. }
\label{fig7}
\end{figure}

Next, we show the influence of multiple couplings ($m>1$) on SR in the network ({\ref{eq1}}). Figure {\ref{fig7}} presents the variation of the response amplitude of the last unit for two values of $\delta$ for few values of $m$. In Fig.~{\ref{fig7}}(a) for $\delta=0.75$ the variation of $Q_{400}$ with $m$ is almost negligible. For $\delta=1$, in Fig.~{\ref{fig7}}(b), the effect of number of couplings is clearly evident. For the entire range of values of $D$ the response amplitude $Q_{400}$ is $> Q_1$. $D_{{\mathrm{max}}}$ is independent of $m$. For $m>3$ we find that $Q_{400}(m) < Q_{400}(m=1)$, however, it is much higher than $Q_1$. We thus conclude that coupling to one or two units is more effective than more number of couplings.

\subsection{Network with periodic signal and noise applied to all the units}
Now, we study the case of the network where all the units are driven by the periodic force and noise and the units are again unidirectionally coupled. The network model is given by
\begin{eqnarray}
 \label{eq3}
  x_{n+1}^{(i)} 
     & = & \frac{r x_n^{(i)} }{ 1 + \left( x_n^{(i)} \right)^b}
               + f \cos \, \omega n + \sqrt{D} \xi_i (n)
               + \frac{\delta \epsilon_i}{M}  \sum_{j=1}^m x_n^{(i-j)} ,  
            \nonumber \\
     &   & \;\; i=1,2,...,N, \;\; i-j \ge 1       
\end{eqnarray}
where $M$ is given by Eq.~({\ref{eq1}}c) and $\epsilon_i=0$ for $i=1$ and $1$ for $i>1$. $\xi_i(n)$ indicates that the units are being under different and independent Gaussian white noise with zero mean and $\langle \xi_i(n) \xi_i(n') \rangle = \delta(n-n')$. 

There are differences in the signal propagation when all the units are driven. Figure {\ref{fig8}} presents the three-dimensional plot of dependence of the response amplitude with the unit number $i$ and the noise intensity $D$ for four fixed values of the coupling strength $\delta$ with $m=1$. Evidently,  each unit exhibits SR with respect to noise intensity for each fixed value of $\delta$. Before discussing on the Fig.~{\ref{fig8} we show in Fig.~{\ref{fig9}} the variation of $D_{{\mathrm{max}}}$ and $Q_{{\mathrm{max}}}$ of various units for $\delta \in [0,1]$.
\begin{figure}[t]
\begin{center}
\epsfig{figure=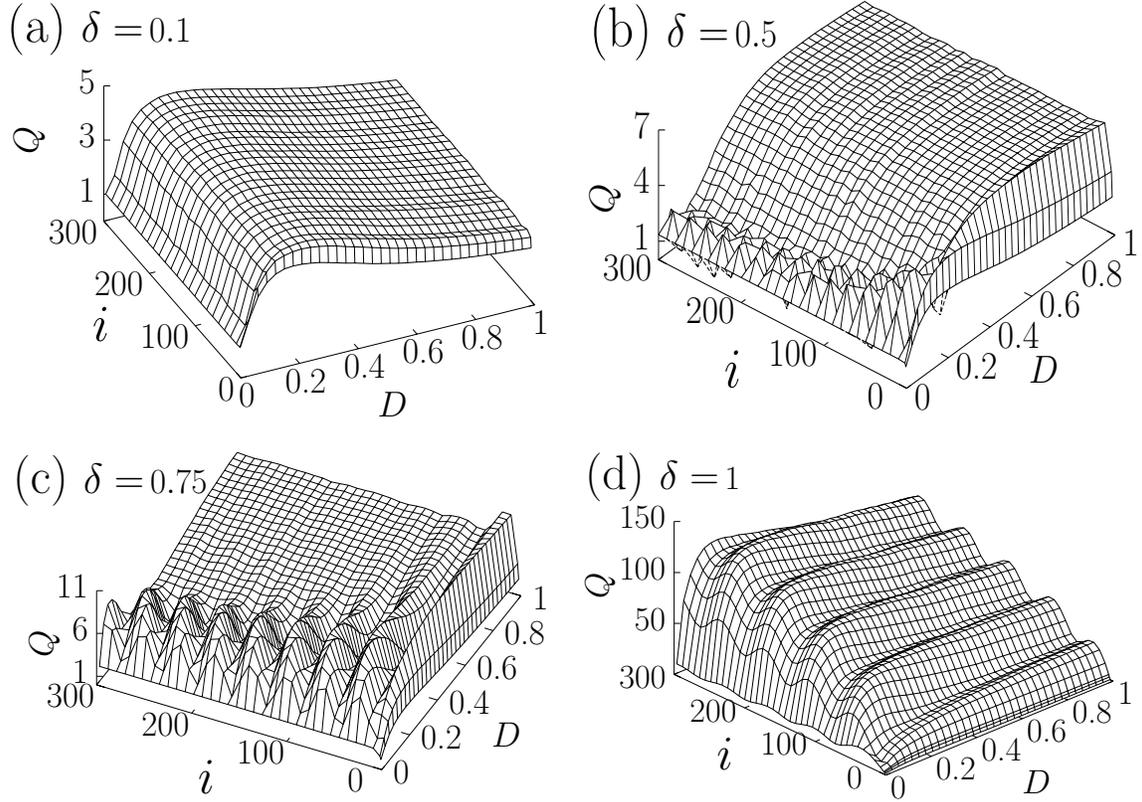, width=0.92\columnwidth}
\end{center}
\caption{Variation of the response amplitude of various units with the noise intensity for four fixed values of the coupling constant $\delta$ of the network system ({\ref{eq3}}) where all the units are driven by the periodic force and noise.  The values of the parameters are $r=2$, $b=2$, $f=0.3$, $\omega=0.1$, $N=400$ and $m=1$.  }
\label{fig8}
\end{figure}
\begin{figure}[!h]
\begin{center}
\epsfig{figure=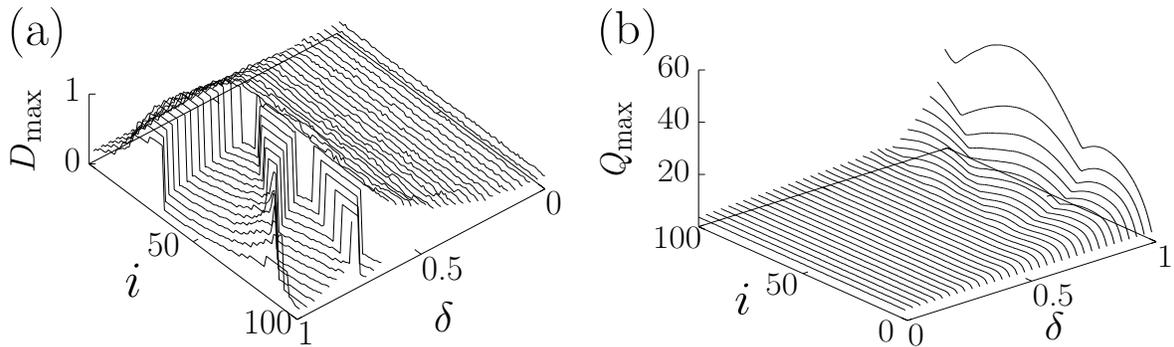, width=0.95\columnwidth}
\end{center}
\caption{Plots of (a)  $D_{{\mathrm{max}}}$ and (b) $Q_{{\mathrm{max}}}$ versus the unit number $i$ and the coupling strength of the network ({\ref{eq3}}) with $m=1$.  }
\label{fig9}
\end{figure}
$D_{{\mathrm{max}}}$ of each unit varies with $\delta$ (except for very small values) and further for a fixed value of $\delta$ its values for various units are not the same. We recall that in the case of the network ({\ref{eq1}}) with the periodic force and noise applied to the first unit only, $D_{{\mathrm{max}}}$ is the same for all the units and independent of $\delta$. In Fig.~{\ref{fig8}}(a) ($\delta=0.1$) the value of $Q_i$ at resonance increases with $i$ and then becomes a constant with $Q_{N,{\mathrm{max}}}>Q_{1,{\mathrm{max}}}$. For $\delta=0.5$, in Fig.~{\ref{fig8}}(b), we observe oscillatory variation of $Q_i$ with the unit number $i$ for small values of $D$. The range of values of $D$ over which $Q_i$ oscillates with $i$ and its period of oscillation increase with increase in $\delta$. The oscillation of the response amplitude pronounces for higher values of $\delta$. Though $Q_i$ and $Q_{{\mathrm{max}}}$ oscillate an interesting result is that for each fixed value of $D$ we find $Q_i(i>1) > Q_1$. The oscillatory variation of $Q_i$ is not found for the network system ({\ref{eq1}}). In Figs.~{\ref{fig4}} and {\ref{fig9}} we observe very high enhancement of response amplitude at resonance for a wide range of values of $\delta$. Our study indicates that driving the first unit alone by the periodic force and noise with unidirectional coupling is a better one compared to driving all the units. 
\section{Stochastic resonance in a network with random coupling}
In this section we consider a network of Bellows map with random one-way coupling with varying connection probability. We study the effect of $25\%$, $50\%$, $75\%$ and $100\%$ of total units coupled (connected) to randomly chosen units on SR. Self-feedback is avoided. All the units are subjected to weak input periodic signal and noise.

The network is governed by the equations 
\begin{eqnarray}
 \label{eq4}
  x_{n+1}^{(i)} 
     & = & \frac{r x_n^{(i)} }{ 1 + \left( x_n^{(i)} \right)^b}
               + f \cos \, \omega n + \sqrt{D} \xi_i (n)
               + \frac{\delta \epsilon_i}{m}  \sum_{j=1}^m x_n^{(M_{ij})} ,  
                 \;\; i=1,2,...,N,       
\end{eqnarray}
where $N$ is the total number of units in the network, $m$ is the number of couplings, $M_{ij}(\ne i)$, $j=1,2,...,m$ are randomly chosen $m$ distinct units and $\epsilon_i=1$ if the $i$th unit is chosen for connectivity otherwise it is $0$. In the numerical simulation we use $r=2$, $b=2$, $f=0.3$, $\omega=0.1$ and $N=400$. 

Because the number of units selected for coupling and the units with which selected units are coupled are chosen randomly we use the average response amplitude $\langle Q \rangle$ to capture SR. We calculate $\langle Q \rangle$ using the following procedure. After every iterations of Eqs.~({\ref{eq4}}) (leaving sufficient transient) we find the average value of $x_n^{(i)}$ given by $\langle x_n \rangle =(1/N) \sum_{i=1}^{N}x_n^{(i)}$. Using these average values over $10^3T$, where $T=2 \pi / \omega$, iterations we compute the response amplitude using the Eqs.~({\ref{eq2}) and denote it as $\langle Q_j \rangle$. It is the response amplitude of, say, $j$th realization of the network topology. We repeat this process for $100$ different realizations of the network topology and then compute the average of $\langle Q_j \rangle$ and call it as average response amplitude $\langle Q \rangle$. 

First we fix $m=1$ in Eqs.~({\ref{eq4}}) and show the effect of size $N$ of the network on $\langle Q \rangle$. Further, we chose $\delta=0.3$. Figure {\ref{fig10}}(a) presents the variation of $\langle Q \rangle$ with the total number of units $N$ for few values of noise intensity. Only $N/4$ units are chosen for connection. Figure {\ref{fig10}}(b) shows the result for all the $N$ units set into connection. In both the cases $\langle Q \rangle$ increases or decreases sharply depending on the value of $D$ with $N$ and then reaches a saturation. 
\begin{figure}[t]
\begin{center}
\epsfig{figure=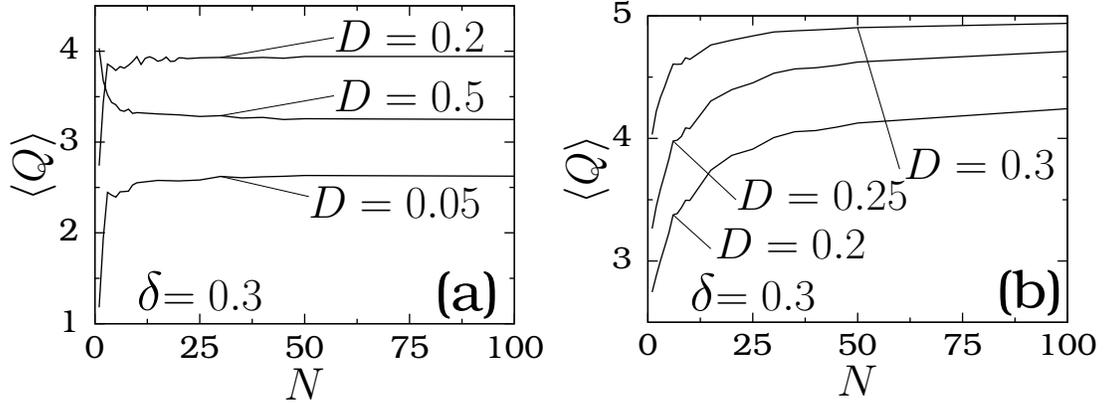, width=0.9\columnwidth}
\end{center}
\caption{$\langle Q \rangle$ versus the total number of units in the network given by Eqs.~({\ref{eq4}}) for three fixed values of noise intensity with $\delta=0.3$. (a) $25\%$ and (b) $100\%$ of the total units are wired to randomly chosen units. The number of coupling is $1$. }
\label{fig10}
\end{figure}
Figure {\ref{fig11}} shows three-dimensional plot of $\langle Q \rangle$ versus $D$ and $\delta$ for $N/4$ and $N$ units are wired. The curves display typical SR character. The effect of $\delta$ on the resonance profile can be clearly seen. Increasing the value of $\delta$ requires increasingly higher $D$ for the maximum response. In addition to the coupling strength $\delta$ the number of random connections also affects the resonance. 
\begin{figure}[t]
\begin{center}
\epsfig{figure=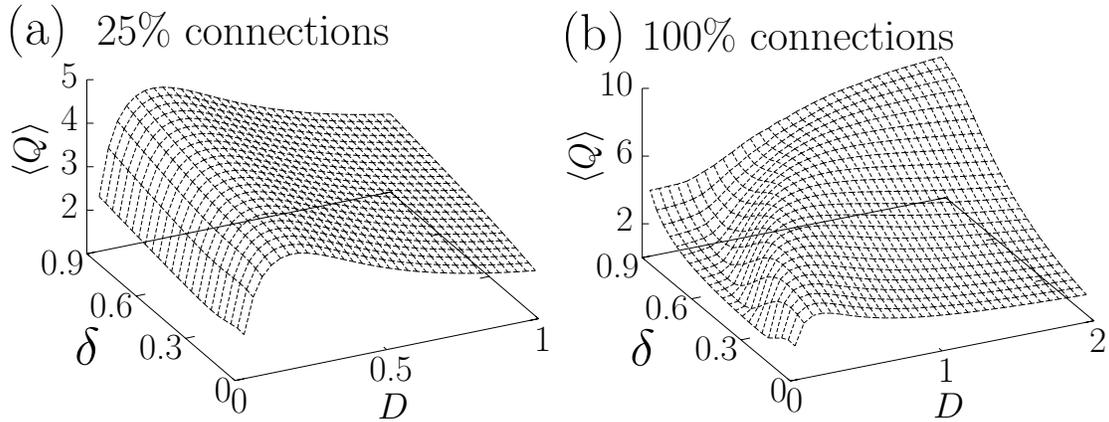, width=0.9\columnwidth}
\end{center}
\caption{$\langle Q \rangle$ as a function of the coupling strength $\delta$ and the noise intensity $D$ for the network ({\ref{eq4}}) with $m=1$. The subplots (a) and (b) correspond to the cases of $25\%$ and $100\%$ of the units coupled to randomly selected units. }
\label{fig11}
\end{figure}

To gain more insight into the dependence of SR on the number of units coupled, we compute $D_{{\mathrm{max}}}$, the value of $D$ at which resonance occurs, and the corresponding value of average response amplitude denoted as $\langle Q \rangle_{{\mathrm{max}}}$ (best average response amplitude). We consider the variation of these quantities with the coupling strength $\delta$ for $N/4$, $N/2$, $3N/4$ and $N$ units are set connected randomly. In all the cases both the quantities are found to vary monotonically with $\delta$. Further, for a fixed value of $\delta$ both $D_{{\mathrm{max}}}$ and $\langle Q \rangle_{{\mathrm{max}}}$ increases nonlinearly (not shown) with the number of units connected. Thus $\langle Q \rangle_{{\mathrm{max}}}$ can be enhanced either by increasing the coupling strength for a fixed number of connections or by increasing the number f connectivity for a fixed value of $\delta$.

Lastly, we show the influence of number of coupling $m$ on $\langle Q \rangle$. When a unit $i$ is selected for coupling then it is coupled to randomly selected $m$ distinct units avoiding self-coupling. Figure {\ref{fig12}} demonstrates the effect of number of couplings on $\langle Q \rangle$ where $\delta=0.25$. For each fixed value of $D$ the average response amplitude increases with the number of couplings, however, the enhancement of it is quite small. Similar effect is found for higher values of $\delta$ also. 

\begin{figure}[t]
\begin{center}
\epsfig{figure=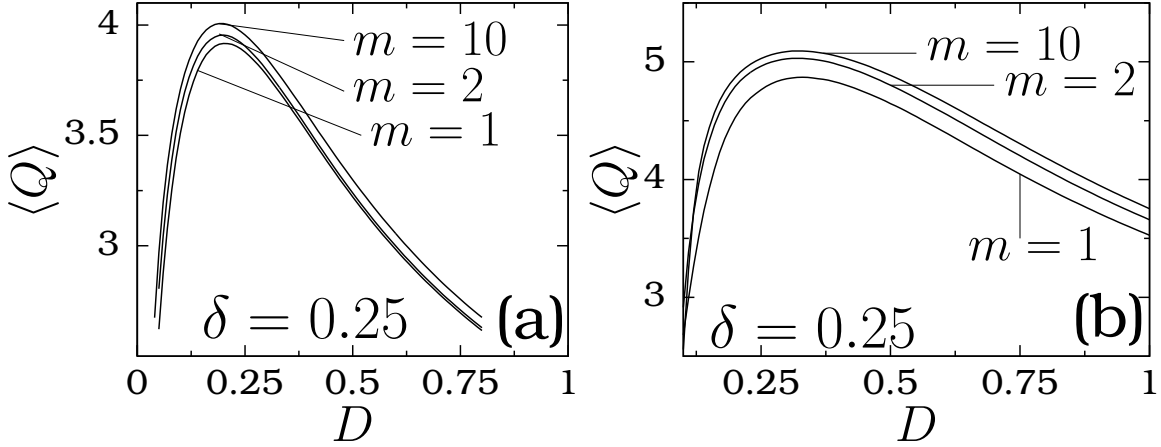, width=0.95\columnwidth}
\end{center}
\caption{Variation of $\langle Q \rangle$ as a function of noise intensity for the system ({\ref{eq4}}) with $\delta=0.25$ and for $1$, $2$ and $10$ couplings. Randomly selected (a) $25\%$ and (b) $100\%$ of the total units are connected to randomly chosen units. }
\label{fig12}
\end{figure}
%
\section{Conclusions}
In the present work we analyzed SR in three types of network systems. In the network ({\ref{eq1}}) periodic signal and noise are added to the first unit only. In the network ({\ref{eq1}}) all the units are driven by the periodic force and noise. The two networks are regular networks as the output of $i$th unit is fed to the ($i+1$)th unit in the case of single coupling. In the network given by Eq.~({\ref{eq4}}) all the units are driven, however, fraction of total number of units are selected randomly for connectivity and are then connected to randomly chosen units. These three networks show certain similarities and difference in the SR phenomenon. In the first network enhanced and undamped propagation of signal at the low-frequency of the input signal occurs above a certain critical value of the coupling strength. In the other two networks enhanced and undamped propagation of signal is realized for the entire range of coupling strength and this is because all the units are driven by the periodic force and noise. Each unit displays only single resonance. Response amplitude varies with the unit number $i$ and attains a saturation. This implies that with certain optimum number of units in a network of the type of connectivities considered in the present work one can realize maximum response. 

In the network ({\ref{eq1}}) the oscillation induced about the two fixed points by the periodic force and noise decreases with increase in the unit number. For the distant units the output becomes a sequence of rectangular pulse and this property is observed for all values of noise intensity. This dynamics is not observed in the other two networks. In the network ({\ref{eq1}}) the value of noise intensity $D_{{\mathrm{max}}}$ is independent of the coupling strength $\delta$ while in the other two networks it varies with $\delta$. Referring to the Figs.~{\ref{fig7}} and {\ref{fig12}} we infer that with single coupling alone we can realize great enhancement of response amplitude. We add that the observation of enhancement of response amplitude across the whole network by applying the input signal and noise to only one unit is greatly useful for weak signal detection and information propagation. 

\newpage
  
\section*{Figure Caption}

\vskip 11pt

\noindent{\bf{FIG.1:}} Examples of one-way coupled networks with $10$ units. The dynamics of the first unit is independent of the other units and this is alone driven by both weak periodic signal and noise. (a) $i$th unit ($i \ne 1$) is linearly coupled to ($i-1$)th unit. The arrow mark indicates that the output of $i$th unit is fed to ($i+1$)th unit only through the linear coupling term. (b) $i$th unit ($i>2$) is linearly coupled to both ($i-1$)th and ($i-2$)th sites while the second site ($i=2)$ is coupled to the first site only.

\vskip 11pt

\noindent{\bf{FIG.2:}} Three-dimensional plots of variation of the response amplitude $Q$ as a function of unit number $i$ and noise intensity $D$ for three values of the coupling strength $\delta$ for the network system given by Eqs.~({\ref{eq1}}). The values of the parameters are $r=2$, $b=2$, $f=0.3$, $\omega=0.1$, $N=400$ and $m=1$. 
 
\vskip 11pt

\noindent{\bf{FIG.3:}} Variation of the response amplitude with the unit number $i$ for six fixed values of the noise intensity $D$ with $\delta=0.75$ and $m=1$. The values of $D$ for the curves $1-6$ are $0.002$, $0.005$, $0.02$, $0.05$, $0.175$ and $0.5$ respectively.

\vskip 11pt

\noindent{\bf{FIG.4:}} $Q_{{\mathrm{max}}}$ of the last unit with the coupling strength $\delta$. The dashed line represents the value of $Q_1$ which is independent of $\delta$.

\vskip 11pt

\noindent{\bf{FIG.5:}} Regions of various ranges of gain factor $G$ in the $(\delta,D$) parameter space of the network system given by Eqs.~(\ref{eq1}) with $m=1$. The motion is unbounded for $\delta>1$.

\vskip 11pt

\noindent{\bf{FIG.6:}} $x_n^{(i)}$ versus $n$ for the units $i=1$, $2$ and $100$ of the network (\ref{eq1}) with $m=1$, $\delta=0.75$ and $D=0.175$ at which the response amplitude becomes a maximum. $x_n^{(i)}$ exhibits periodic switching between the regions $x>0$ and $x<0$. 

\vskip 11pt

\noindent{\bf{FIG.7:}} The effect of multiple coupling on the response amplitude of the last unit of the network system given by Eqs.~({\ref{eq1}}) for two values of the coupling constant $\delta$. In the subplot (b) the number marked to each curve is the value of $m$ (number of couplings). The dashed curve is the response amplitude $Q_1$.

\vskip 11pt
\newpage

\noindent{\bf{FIG.8:}} Variation of the response amplitude of various units with the noise intensity for four fixed values of the coupling constant $\delta$ of the network system ({\ref{eq3}}) where all the units are driven by the periodic force and noise. The values of the parameters are $r=2$, $b=2$, $f=0.3$, $\omega=0.1$, $N=400$ and $m=1$. 

\vskip 11pt

\noindent{\bf{FIG.9:}} Plots of (a) $D_{{\mathrm{max}}}$ and (b) $Q_{{\mathrm{max}}}$ versus the unit number $i$ and the coupling strength of the network ({\ref{eq3}}) with $m=1$.

\vskip 11pt

\noindent{\bf{FIG.10:}} $\langle Q \rangle$ versus the total number of units in the network given by Eqs.~({\ref{eq4}}) for three fixed values of noise intensity with $\delta=0.3$. (a) $25\%$ and (b) $100\%$ of the total units are wired to randomly chosen units. The number of coupling is $1$. 

\vskip 11pt

\noindent{\bf{FIG.11:}} $\langle Q \rangle$ as a function of the coupling strength $\delta$ and the noise intensity $D$ for the network ({\ref{eq4}}) with $m=1$. The subplots (a) and (b) correspond to the cases of $25\%$ and $100\%$ of the units coupled to randomly selected units. 

\vskip 11pt

\noindent{\bf{FIG.12:}} Variation of $\langle Q \rangle$ as a function of noise intensity for the system ({\ref{eq4}}) with $\delta=0.25$ and for $1$, $2$ and $10$ couplings. Randomly selected (a) $25\%$ and (b) $100\%$ of the total units are connected to randomly chosen units.

\end{document}